\documentstyle[twocolumn,amssymb,aps,psfig,float]{revtex}
\begin{document}
\tolerance 50000
%%%%\preprint{
%%%%\begin{minipage}[t]{1.8in}
%%%%\hfill LPQTH-94/?
%%%%\end{minipage}
%%%%}  

\draft
\twocolumn[\hsize\textwidth\columnwidth\hsize\csname @twocolumnfalse\endcsname
\title{Double butterfly spectrum for two interacting particles in the Harper 
model}

\author {Armelle Barelli, Jean Bellissard, Philippe Jacquod$^{a}$ and Dima L.
Shepelyansky$^{b}$}
\address {Laboratoire de Physique Quantique, UMR 5626 du CNRS, Universit\'e Paul
Sabatier, F-31062 Toulouse Cedex, France\\ 
$^{a}$ Institut de Physique, Universit\'e de Neuch\^atel, CH-2000 Neuch\^atel,
Conf\'ed\'eration Helv\'etique}
%\date{\today}
\date{15 September, 1996}
\maketitle

\begin{abstract}
\begin{center}
\parbox{14cm}{
We study the effect of interparticle interaction $U$ 
on the spectrum of the Harper
model and show that it leads to a pure-point component 
arising from the multifractal
spectrum of non interacting problem. Our numerical studies allow to
understand the global structure of the spectrum. Analytical approach developed
permits to understand the origin of localized states in the limit of strong 
interaction $U$ and fine spectral structure for small $U$.
}
\end{center}
\end{abstract}

\pacs{
\hspace{1.9cm}
}
\vskip2pc]

Recently a great deal of attention has been devoted to the investigation of
incommensurate systems exhibiting singular continuous spectrum with many
interesting multifractal properties 
(see e.g. \cite{Geisel,Guarn,Piech}). Among the
physical models, one of the most popular is the Harper model of electrons on a 
two-dimensional square lattice in the presence of a perpendicular magnetic 
field \cite{Harper,Hofst}. This system can be reduced to the study of a rather
simple model of particle 
dynamics on a one-dimensional quasiperiodic lattice. The energy spectrum
exhibits multifractal properties and the band spectrum for rational
values of magnetic flux looks like a butterfly. In spite of the academic
character of such a model, experiments have been performed during the last ten
years exhibiting this multifractal butterfly structure. One of
the first among them has been performed in 1985 using superconducting networks
\cite{Pannetier} and more recently experiments with superlattices also allowed
to observe the fist hierarchical steps of multifractal butterfly structure
\cite{Gerhardts}.

The deep understanding of such an intricate spectral structure attracted
interest of mathematicians and physicists who developed new approaches for its
investigation such as non commutative geometry \cite{Bel}, pseudo-differential
operators \cite{HeSj}, functional analysis \cite{Last}, renormalization
group approach \cite{Thouless,Wilki}, thermodynamical formalism \cite{KH}.   
All these tools allowed to study the problem on rigorous mathematical ground and
to understand the properties of eigenstates. For example using the duality
between momentum and spatial coordinate \cite{Aubry}, it is possible to prove
rigorously the existence of localized or delocalized states
\cite{BeBa,GuarBorgo}. It was also found that quantum systems which are chaotic
in the classical limit may have quite unusual properties in the presence of
underlying quasiperiodic structure \cite{LimaShepe,Geisel,KH}.

All the works mentioned above were done for one particle dynamics. However even
from the physics of the original Harper model, it is clear that the interaction 
between electrons on the square lattice in the presence of magnetic flux plays 
an important r\^ole. Therefore it is natural to address the question of the
influence of interaction on multifractal spectrum. The most
simple example of such a case is an interaction
between two particles. Recently it has been found that in the case of random
potential even such simple model has a number of unexpected properties
\cite{TIP}. For example repulsive/attractive short range interaction leads
to appearance of effective pair states in which 
two particles propagate together
on a distance much larger than the one-particle localization length without
interaction. Surprinsingly the first numerical studies of interaction effect in
a quasiperiodic potential showed an opposite tendency \cite{Shepe96}. Namely,
repulsive/attractive interaction leads to the 
appearance of localized states while
in the absence of interaction multifractal spectrum generated quasidiffusive
spreading of wave packets on the lattice. However, the numerical approach used
in \cite{Shepe96} allowed to study only the wave packet evolution while the 
structure of the spectrum itself was not directly accessible. Therefore to
understand the spectral structure and the nature of eigenstates we
performed numerical simulations by direct diagonalisation 
based upon Lanczos algorithm. 

As a basic model for our investigations we consider the model of two 
interacting particles (TIP) in the Harper problem described by the following
eigenvalues equation 
%1
\begin{equation}
\begin{array}{c}
(2\lambda \cos(\gamma n_1+\beta_1)+
2\lambda \cos(\gamma n_2+\beta_2)+U\delta_{n_1, n_2})\varphi_{n_1, n_2} + \\
\varphi_{n_{1}+1, n_2}+\varphi_{n_{1}-1, n_2}+ 
\varphi_{n_{1}, n_{2}+1}+\varphi_{n_{1}, n_{2}-1}=E\varphi_{n_{1}, n_{2}}
\end{array}
\end{equation}
where the parameter $ \gamma $ characterizes the quasiperiodic
lattice for the one-particle problem. Without interaction, each particle moves
in quasiperiodic Harper potential and  $ \gamma /2\pi =\phi /\phi_0 =\alpha $ is the
ratio between the magnetic flux within one unit cell of the square lattice and
the flux quantum $ \phi_0=h/e $. The parameter $ \alpha $ plays the role of an
effective Planck's constant so that $ \alpha\mapsto 0 $ corresponds to
the semiclassical limit. The two parameters $ \beta_{1,2} $ are related
to the quasimomentum components in the non interacting problem. The parameter $ \lambda $ characterizes the
strength of the quasiperiodic potential and for the case of electrons on a
square lattice $ \lambda =1 $ \cite{Hofst}. However from mathematical point of
view it is also interesting to study the different regimes with $ \lambda <1 $ and 
$ \lambda >1 $. Strong analytical and numerical evidence has been given that the
spectrum is pure point and the states are localized when $ \lambda >1 $ while
for $ \lambda <1 $ the spectrum is continuous with extended eigenstates
\cite{Aubry,Last,Geisel,Wilkinson}. The strength of the short range on-site 
interaction is characterized by $ U $. We concentrate our investigations
on the case $\lambda=1$, $\beta_{1,2}=\beta$ 
when for $U=0$ the spectrum is multifractal 
for irrational values of $\gamma/2\pi$. We consider only 
the part of the spectrum corresponding to the symmetric 
TIP states since antisymmetric
configuration is not affected by on-site interaction.

In the absence of interaction, the corresponding two particle spectrum results
of the superposition of two one-particle spectra of the Harper model and is
shown in Fig. 1 (a). Comparing with the one-particle spectrum (Hofstadter's
butterfly), we can remark that the spectrum becomes much more dense near the
centers of the bands and subbands but still the gaps in the spectrum survive on
all energy scales.

\vbox to 180bp {\vfil
\centerline{\hbox to 320bp 
{
%\special{psfile=fig1a.ps 
%hsize=320 vsize=180
%hscale=34 vscale=34
%angle=90}\hfil
}}}

%\vbox to 180bp {\vfil
%\centerline{\hbox to 360bp 
%\special{psfile=fig1b.eps
%hscale=55.675 vscale=47.35 hoffset=-46 voffset=-210
%angle=00}
\vglue 65truemm

\noindent {\it {\bf Fig. 1} : Spectrum of two particle Harper problem (a : up), 
with $ U=0 $
obtained for rational values of $ \gamma /2\pi =\alpha = p/q $ with $ q \leq 19$;
(b : down) with $ U=1 $ and $ q \leq 23$.}

When increasing the strength of the interaction $ U $, the spectrum is splitted
into two butterflies which are slightly shifted one respect to the other. 
However
one of them remains almost at the same place corresponding to the non
interacting case of Fig. 1 (a). The shifted butterfly moves to the right since
the repulsive interaction $U>0$ gives global increase of energy. A typical 
case $U=1$ of double butterfly spectrum is presented in Fig. 1 (b).

The main features which can be immediately observed in this figure are
the smoothness of the edge of the shifted butterfly, 
the less dense character of
its spectrum and the filling of some internal energy gaps (see for example near
$ \alpha =0.6 $ and $ E=-1.5 $). However, the gaps in the spectrum still exist
on all scales. 

The shift
of one butterfly and almost unchanged form for the other at moderate
values of interaction $ U $ can be understood in the following simple way. 
For that we choose small values of flux $ \alpha\ll
1 $ and use the perturbation theory in $ U $ 
on the basis of harmonic oscillator
functions to get analytical expressions for the Landau sublevels at the 
spectrum edge. Without interaction, the band edge is given by $ E_{\pm }(\alpha )
=\pm 8\mp 4\pi\alpha (m_1+m_2+1)\pm \pi ^2\alpha ^2 \left( 2+(2m_1+1)^2+(2m_2+1)
^2\right) /4 + {\rm O}(\alpha ^3) $, which is superposition of two Hofstadter
butterflies in semiclassical regime \cite{RaBe}. The integers $ m_1,m_2 $
are the Landau quantum numbers for oscillator states near the bottoms of
potential minima. If two particles are located in different minima, the
interaction between them is negligibly small and the energy levels are not
shifted by $ U $. These energy states correspond to non shifted butterfly with
dense spectrum since there are many states when 
TIP are separated from
each other. If TIP are located in the same potential minimum, the
interaction gives energy shift which in the first order of perturbation theory
is $ \Delta E_{\pm } = U\sqrt{\alpha } $ for $ m_{1,2}=0 $ and $ m_{1,2}=(0;1)$
being in good agreement with numerical data for $U<1$ as can be seen
on Fig. 2 (a). 
This shows that the shifted butterfly corresponds to the case when
the two particles are located near each other. The density of such states is
smaller than in the case when particles are far from each other and that is 
why the shifted butterfly is less dense. 

\includegraphics{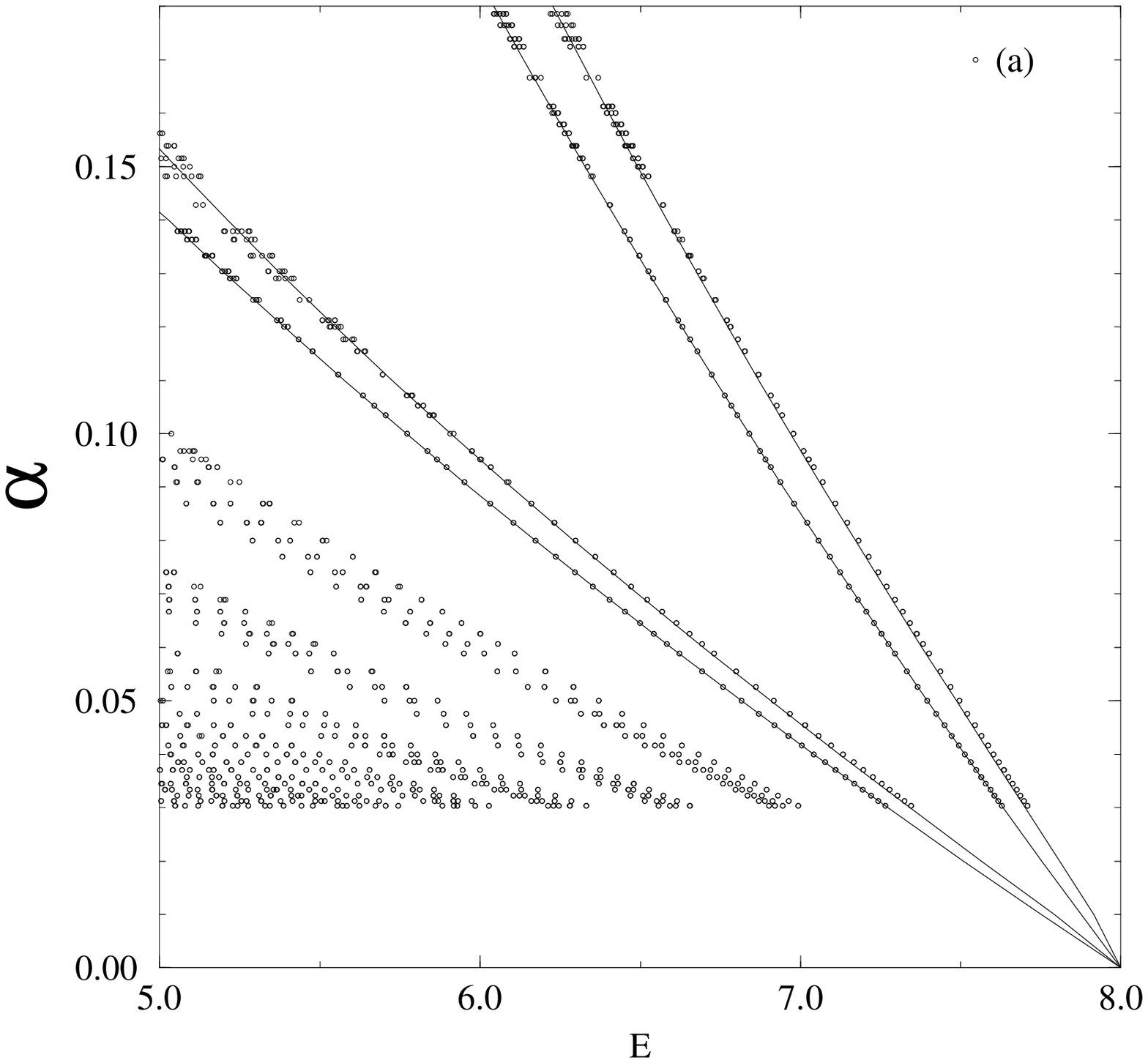}
\includegraphics{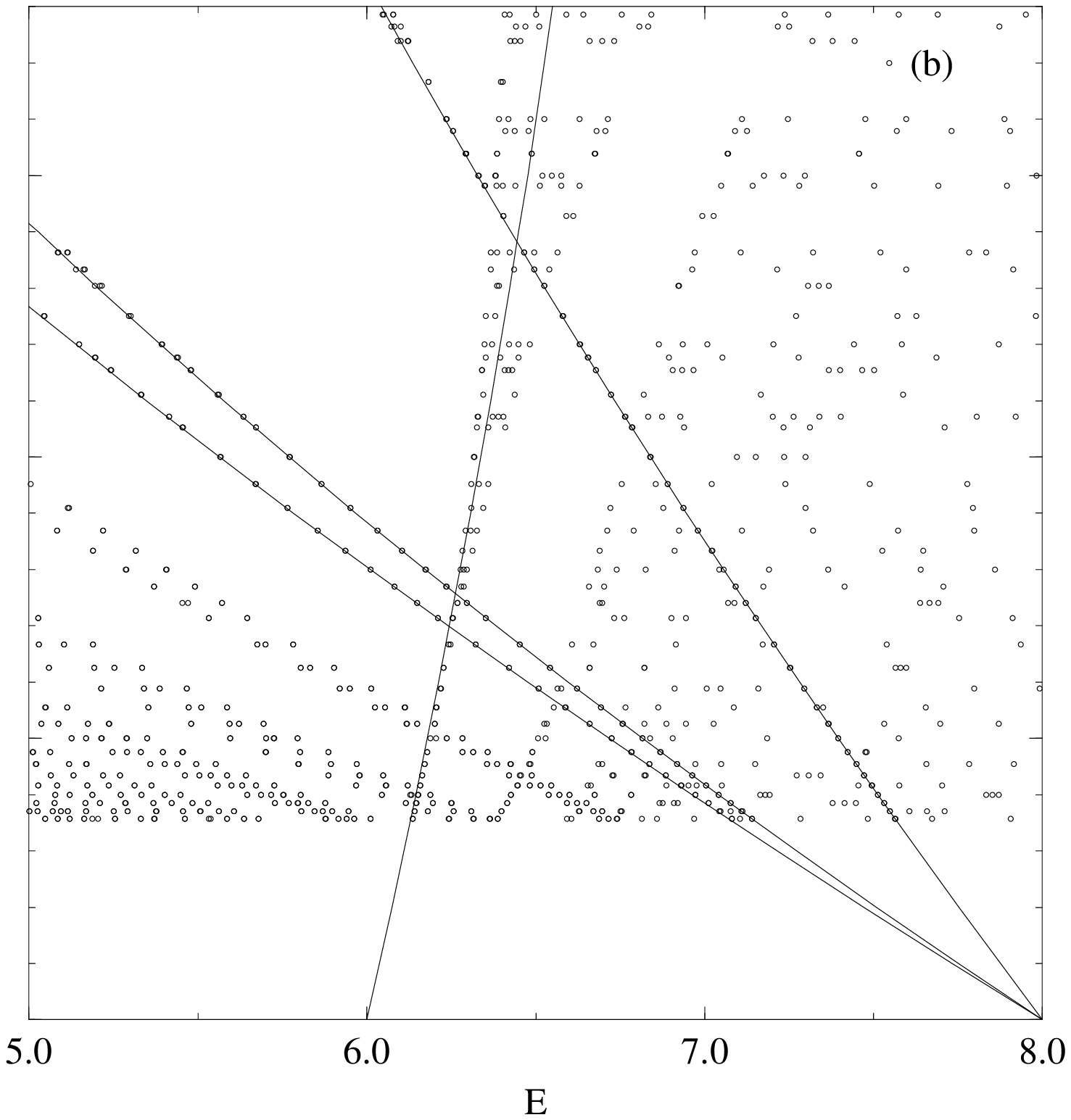}
\vglue 45truemm

\noindent {\it {\bf Fig. 2} : Energy band edges (a) U=0.4, dots are
numerical data and solid curves are perturbation theory results (see text); 
(b) U=10, 
dots are data from Fig. 4 and solid curves are given by theory described in the
text.}

Direct analysis of eigenstates for irrational flux values 
(which are approximated by a continuous fraction expansion) also shows that the
states in the shifted part correspond to the situation where two particles stay
near each other. However, contrary to the TIP in a random
potential, the particles here cannot propagate together and stay exponentially 
localized near the origin as it can be seen with the typical 3-D plot of Fig. 3. 
   
%\special{psfile=fig3a.eps 
%hoffset=-70 voffset=-260
%hscale=50 vscale=50
%angle=00}
%\special{psfile=fig3b.eps 
%hoffset=50 voffset=-260
%hscale=50 vscale=50
%angle=00}
\vglue 5truecm

\noindent {\it {\bf Fig. 3} : Semilog plot of
$W_{n_1,n_2}=\vert\phi_{n_1,n_2}\vert ^2 $ 
for localized ($ E=-1.3376, -10 \leq \ln W \leq -1, \xi =5.9, \xi _0=193 $ (a :
left)) 
and  delocalized ($ E=-1.7368, -10 \leq \ln W \leq -3, \xi =214, \xi _0=12.5 $
(b : right)) eigenstates at $U=1, \alpha = 34/55 , \beta =\sqrt{2} $.}

\vspace{3mm}

We also investigated the structure of eigenstates 
in the more dense part of the
spectrum (non shifted butterfly). In that case the eigenstates are delocalized
and quite similar to those corresponding to the non interacting case. 
Here the two
particles mainly spread quasidiffusively along the quasiperiodic lattice and
interaction is not important for them. This structure of localized and
delocalized eigenstates is in agreement with the numerical study of wave packet
dynamics performed in \cite{Shepe96}.

The properties of eigenstates can also be analyzed with the help of the inverse
participation ratio (IPR) $ \xi =( \sum _{n_{1,2}}W_{n_1,n_2}^2) ^{-1} $.
Its value for different energies is shown in Fig. 4 for $ \alpha =34/55 $. In
agreement with the above discussion, the localized states with small $ \xi $ correspond to
the part of the shifted butterfly with less dense spectrum while the unshifted
butterfly is associated to large $ \xi $ with delocalized states. It is
interesting to determine the IPR $ \xi _0 $ in the non interacting eigenstates 
basis. Such approach has been quite useful for TIP in a random potential
\cite{Shepe95}. It is interesting to note that the situation for TIP in the 
Harper model is quite different. Namely, the delocalized states have very small
value of $ \xi _0 $ while the localized ones are delocalized in the non
interacting eigenstates basis and have very large $ \xi _0 $
(see Fig.3). This result once
more shows that delocalized states correspond to almost non interacting particle 
propagation while localized states appear only due to interaction which can be
even repulsive (Fig.3a).

With further increase of $ U $ the shifted butterfly goes on moving to the right
and becomes more and more deformed. Starting from interaction strength $ U\geq
10 $, this butterfly is transformed into a spectral band with width two times
smaller than the original spectrum at $ U=0 $. The center of this band is 
located at energy $ E\approx U $. The typical example of global spectrum is
shown in Fig. 5.

%\special{psfile=fig4.eps 
%hoffset=00 voffset=-210
%hscale=45 vscale=45
%angle=00}
\vglue 65truemm

\noindent {\it {\bf Fig. 4} : Inverse participation ratios $ \xi $ 
vs eigenenergies $ E $ shown at $ \xi = 2 $; 
$ U=1, \alpha =34/55, 0 \leq \beta < 2 \pi $.}

%\special{psfile=fig5.eps 
%angle=00 hscale=50.1 vscale=42.615 voffset=-210 hoffset=-20}
\vglue 65truemm

\noindent {\it {\bf Fig. 5} : Same as in Fig. 1 with 
$ U=10 $ and $ q \leq 28$.}
  
The physical reason for the appearance of such separated spectral band can be
understood in the following way. For strong $ U $, there are states for which 
TIP are localized on the same site so that $ n_{1,2}=n $. According to (1) the
energy of the states is $ E_n=4\lambda\cos (\gamma n+\beta ) +U $. 
The transition between these states
can be obtained with first order perturbation theory in $ 1/U $ which gives the
effective eigenvalue equation :
%2
\begin{equation}
\left( 4\lambda\cos (\gamma n+\beta ) +U\right)\phi _{n}+V_{\rm eff}\left(\phi
_{n+1}+\phi _{n-1}\right) =E\phi _{n}
\end{equation}
Here $ V_{\rm eff} $ is the hopping between such states due to virtual
transitions via states with $ n_1\neq n_2 $. For $ U\gg 1 $, the energy
difference between diagonal and off-diagonal states is very large and therefore
$ V_{\rm eff}\sim 1/U $. The equation for diagonal eigenstates has the form of
Harper equation with $ \lambda _{\rm eff}=2\lambda /V_{\rm eff}\gg 1 $. Due to
that these states are exponentially localized so that particles stay near the 
origin. In some sense, the interaction renormalizes the constant $
\lambda\mapsto\lambda _{\rm eff} $
in the Harper equation for pair of particles. For strong $ U $, 
the renormalized
$ \lambda _{\rm eff} $ is much larger than 1 that, according to the Aubry
duality \cite{Aubry}, leads to localization of TIP pairs in
quasiperiodic potential. Our conjecture is that in a sense $ \lambda _{\rm eff} 
$ remains larger than 1 even for moderate values of $ U \sim 1 $. 
In a sense interaction breaks Aubry duality leading to 
appearence of localized TIP phase.
However more rigorous
analytical confirmations of this conjecture are desirable especially keeping in
mind that in a random potential the interaction with $U \sim 1$
leads to delocalization of TIP pair states. 
The accurate expressions for the TIP energy edges of 
shifted spectral band can be
found using semiclassical analysis at small flux 
values by methods developed in 
\cite{Bel,BaFl}. The details of computations will be given
elsewhere \cite{BaBeJaSh}. For the case of Fig. 2 (b), 
they give $ E=6.0 + 0.59*
2\pi\alpha + {\rm O}(\alpha ^2 ) $ that is in good agreement with numerical
data (Fig. 2 (b)). 

For the part of the spectrum represented by unshifted butterfly 
at $U \gg 1$, the
eigenstates become more and more similar to asymmetric TIP configuration i.e. 
$ \phi _{n_1,n_2}={\rm sign}(n_1-n_2)\left(\chi ^{(1)}_{n_1}\chi ^{(2)}_{n_2}-
\chi ^{(1)}_{n_2}\chi ^{(2)}_{n_1}\right) /\sqrt{2} $,
where $\chi$-s are one-particle eigenfunctions. Due to that, the
effective interaction becomes quite small and 
the unshifted butterfly at large $
U $ (TIP are in different wells) looks very similar to the one at $ U=0 $. 
The main difference is the
splitting of Landau sublevels which appears due to effective small interaction
between particles located in the same well. According to the expression for $
\phi _{n_1,n_2} $, such splitting can take place only when Landau quantum
numbers are different ($ m_1\neq m_2 $) so that $ \chi ^{(1)}\neq \chi ^{(2)} $.
As the result the first sublevel with $ m_{1,2}=0 $ is not splitted. For non
interacting part, the edges are given by the same $ E_{\pm }(\alpha ) $ as for $
U=0 $ (see above) while for interacting case, the additional shift is $ \delta
E(\alpha )=-8\pi\alpha /(U+4) $ (see \cite{BaBeJaSh}). These analytical
expressions are in good agreement with numerical results as shown in Fig. 2 (b).

In summary, {\it 20 years after} \cite{Hofst} our investigations of spectra 
and eigenstates for TIP in the Harper
model (1) show that repulsive/attractive interaction leads to appearance of 
localized states. Our conjecture is 
that due to Aubry duality breaking a localized TIP pair phase appears at
arbitrary small interaction strength. 
At the same time we expect that this breaking is absent 
for TIP on the 2d-lattice with magnetic flux.
However, the later model requires separate investigations
\cite{BaBeJaSh}.

This work is supported in part by the Fonds National Suisse de la
Recherche.

\end{document}